\documentstyle[prl,aps,preprint,tighten]{revtex}
\begin{document}
\draft
\title{Spectral properties of Luther-Emery systems}
\author{Johannes Voit}
\address{Bayreuther Institut f\"{u}r Makromolek\"{u}lforschung (BIMF) and \\
Theoretische Physik 1, Universit\"{a}t Bayreuth, D-95440 Bayreuth (Germany)}
\date{\today}
\maketitle
\begin{abstract}
We calculate the spectral  function of the Luther-Emery model which
describes one-dimensional fermions with gapless charge and gapped spin
degrees of freedom. We find a true singularity with interaction dependent
exponents on the gapped spin dispersion and a finite maximum depending
on the magnitude of the spin gap, on a shifted charge dispersion. We apply 
these results to photoemission experiments on charge density wave systems
and discuss the spectral properties of a one-dimensional Mott insulator.
\end{abstract}

\narrowtext
Non-Fermi liquid behavior in correlated fermion systems is an exciting topic
of current research. One-dimensional (1D) metals are a paradigmatic example
of non-Fermi liquids: their low-energy excitations are not quasi-particles but
rather collective charge and spin density fluctuations which obey each to 
their proper dynamics \cite{myrev}. The key features of these 
``Luttinger liquids'' 
\cite{Haldane} clearly show up in the single-particle spectral function
\begin{equation}
\label{specdef}
\rho(q, \omega) = - \pi^{-1} {\rm Im} \: G(k_F+q, \mu + \omega)
\end{equation}
which can be measured in a photoemsission experiment: (i) absence of fermionic
quasi-particles, (ii) anomalous dimensions of operators producing correlation
functions with non-universal power-laws,  (iii) charge-spin separation
\cite{specll}. [In Eq.~(\ref{specdef}), $G$ is the electronic Green's function,
$k_F$ the Fermi wave number, and $\mu$ the chemical potential.] Responsible
is the electron-electron interaction which is marginal
in one dimension and therefore transfers nonvanishing momentum in scattering
processes at all energy scales, and the nesting properties of the 1D Fermi
surface which allow for the emergence of Peierls $2k_F$ charge and spin 
density fluctuations which then interfere with Cooper-type superconducting
fluctuations. 

In a Luttinger liquid, both the charge and the spin excitations are gapless.
There are, however, other possibilities
for interacting 1D fermions: when backscattering of electrons
with momentum transfer $\pm 2k_F$ becomes relevant (often a consequence of
electron-phonon coupling), a gap in the spin
excitation spectrum opens, while for commensurate band fillings, Umklapp
process may create a charge gap. The other degree of 
freedom would remain gapless. Systems in these classes would be dominated by 
singlet superconducting (SS) \cite{zkl} or charge density wave (CDW) 
correlations \cite{veuill,dardel}
or be 1D Mott insulators \cite{liebwu,preuss} 
-- problems of high experimental and 
theoretical interest. While there is a rather complete picture of the 
properties of
Luttinger liquids \cite{myrev}, much less is known for systems
with both gapless and gapped degrees of freedom. This is particularly
true for dynamical correlation functions such as the spectral function,
Eq.~(\ref{specdef}), which give direct information on the nature and
dynamics of the elementary excitations. There is a general belief that
the opening of a gap affects the system for frequencies smaller than
this gap while the behavior of the ungapped system is essentially recovered
at larger frequency scales. It is the purpose of this Letter
to discuss the spectral function of a model with gapless and gapped 
degrees of freedom, to compare to
the above ``naive'' prediction, and to comment
on recent photoemission experiments on quasi-1D CDW systems where this model
could be relevant.

The generic model describing this situation 
\begin{eqnarray}
\label{ham}
H & = & H_{0}^{(\rho)} + H_{0}^{(\sigma)} + H_{1\perp}^{(\sigma)} \\
\label{hlut}
H_0^{(\nu)} & = & \frac{1}{2 \pi} \sum_{\nu= \rho,\sigma} 
\int \! dx \left\{ v_{\nu} K_{\nu} \: \pi^2
\Pi_{\nu}^2(x) + \frac{v_{\nu}}{K_{\nu}} \left( \frac{\partial 
\Phi_{\nu}(x)}{\partial x} \right)^2 \right\} \;\;\;, \\
\label{hperp}
H_{1\perp} & = & \frac{2 g_{1\perp}}{(2 \pi \alpha)^2} \int \! dx
\cos \left[ \sqrt{8} \Phi_{\sigma}(x) \right] 
\end{eqnarray}
has been solved by Luther
and Emery \cite{luthem}. Here, $H_0$ describes harmonic charge ($\nu = \rho$)
and spin ($\nu = \sigma$) density fluctuations through the bosonic phase
fields $\Phi_{\nu}(x)$ and their canonically conjugate momenta $\Pi_{\nu}(x)$.
Their dispersions are gapless $\omega_{\nu}(q) = v_{\nu} |q|$ with 
velocities $v_{\nu}$, and $H_0$ contains, in addition, stiffness constants 
$K_{\nu}$. The backscattering Hamiltonian $H_{1\perp}$ is, for $K_{\sigma}-1$
small enough compared to $|g_{1\perp}|$, a
relevant perturbation and opens a gap $\Delta_{\sigma}$ in the spin excitation
spectrum. The Umklapp Hamiltonian for a half-filled band is obtained by
simply replacing spin by charge in Eq.~(\ref{hperp}). Luther and Emery have
shown that for the special value $K_{\sigma}=1/2$, the interaction Hamiltonian
(\ref{hperp}) can be represented as a bilinear in spinless fermions, and
diagonalized. The resulting spectrum $\varepsilon_{\sigma}(q) = \pm
\sqrt{v_{\sigma}^2 q^2 + \Delta_{\sigma}^2} $ shows a gap $\Delta_{\sigma}$
at the Fermi level. The possibility of exactly 
calculating correlation functions for this model is severely limited by the
absence of any practical relation between the physical fermions and the
spinless pseudofermions emerging from the Luther-Emery solution \cite{mori}.
Many problems map onto this model at low energies.

Here, I compute the single-particle spectral function $\rho(q,\omega)$, 
Eq.~(\ref{specdef}), for the Luther-Emery model. Other correlation functions
may be obtained along the same lines but less experiments are available
that one could possibly compare to. The charge-spin separation manifest
in the Hamiltonian (\ref{ham}) allows to represent $\rho(q,\omega)$ as a
convolution 
\begin{eqnarray}
\label{specconv}
\rho(q,\omega) & = &  (2 \pi)^{-2} \int_{-\infty}^{\infty} \! d {q'} \:
d {\omega}' \: \left[ g_{\rho}({q}',{\omega}') g_{\sigma}(q-{q}', \omega
-\omega') + (q \rightarrow - q \;, \; \omega \rightarrow - \omega ) \right]
\end{eqnarray}
of certain charge and spin correlation functions
\begin{equation}
\label{corf}
g_{\nu}(x,t) = \langle \Psi^{(\nu)}_{rx}(xt) \Psi^{(\nu)\dag}_{rs}(00) \rangle
\;\;.
\end{equation}
The notation $\Psi^{(\nu)}$ indicates that only the $\nu$-part of the 
boson representation of $\Psi$ is to be taken. 
The charge part is easy (we only display the leading $\omega$- and 
$q$-dependence)
\begin{eqnarray}
g_{\rho}(q,\omega) & \sim &  \Theta(\omega
- v_{\rho} q) \Theta(\omega + v_{\rho} q)
(\omega - v_{\rho}q)^{\gamma_{\rho}-1} 
(\omega + v_{\rho} q)^{\gamma_{\rho} -1/2}  \hspace{0.5cm} (K_{\rho} \neq 1) \\
\label{chargefree}
& \sim & \frac{\Theta(\omega + v_{\rho} q)}{\sqrt{\omega + v_{\rho} q}}
\delta(\omega - v_{\rho} q) \hspace{0.5cm} (K_{\rho} = 1) \; . 
\end{eqnarray}
Using a similar expression for the spins, one can reproduce in detail the
spectral functions of the Luttinger model calculated elsewhere directly
\cite{specll}. Notice that the divergences are stronger than for 
a spinless Luttinger model ensuring that singularities remain  after
performing the convolution integrals.

The determination of the spin correlation function is more involved because
it has no simple representation in terms of the Luther-Emery pseudofermions. 
In real space, I take
\begin{equation}
\label{spreal}
g_{\sigma}(x,t) \sim 
\exp \left( - \Delta_{\sigma} \sqrt{x^2 - v_{\sigma}^2 t^2} 
/ v_{\sigma} \right) / \sqrt{\alpha + i (v_{\sigma}t - x)} \; .
\end{equation}
This form (i) reduces to the correct Luttinger form for vanishing gap.
(ii) From the equivalence of the Luther-Emery model to a classical 2D 
Coulomb gas, and Debye screening of the charges above the Kosterlitz-Thouless
temperature, one deduces the exponential factor \cite{chui}. 
(iii) Gul\'{a}csi has
calculated explicitly the $t=0$-Green function of a 1D Mott insulator
\cite{miklos}: the form (\ref{spreal}) is obtained from his results by
Lorentz-transforming the exponential term as in a massive Dirac theory; 
the power-law prefactor would correspond to a massless theory which is
expected to govern the short space/time ($|x|, v_{\sigma}|t| \ll v_{\sigma} /
\Delta_{\sigma}$) behavior. Fourier transformation then gives
\begin{equation}
\label{spksp}
g_{\sigma}(q,\omega) \sim \left( 1 +
\frac{v_{\sigma} q}{\sqrt{v_{\sigma}^2 q^2 + \Delta_{\sigma}^2}} \right) \:
\frac{\Theta(\omega + v_{\sigma} q)}{\sqrt{\omega + v_{\sigma} q}} \:
\delta(\omega - \sqrt{v_{\sigma}^2 q^2 + \Delta_{\sigma}^2}) 
\end{equation}
This result can then be inserted into the convolution formula (\ref{specconv})
and evaluated. 

What could we expect from our knowledge of the Luttinger liquid \cite{specll}?
There the singularities at $\omega = v_{\rho (\sigma)} q$ 
arise from processes where the charge (spin) contributes all of the electrons'
momentum $q$ and the spin (charge) none. The same argument applied to the
Luther-Emery model predicts singularities at the renormalized spin dispersion
$\varepsilon_{\sigma}(q)$ and at a shifted charge dispersion 
$\varepsilon_{\rho}(q) = v_{\rho} q + \Delta_{\sigma}$ 
(inset in Fig.\ 1). 
The result of the calculation is shown schematically in Fig.\ 1
for $q > 0$ and the (realistic) case $v_{\rho} > v_{\sigma}$. 
There are indeed features at these 
frequencies. At $\varepsilon_{\sigma}(q)$, there is a true
singularity $[\omega - \varepsilon_{\sigma}(q)]^{\alpha-1/2}$ 
as in the Luttinger model (however,
here $\alpha$ is \em defined \rm as $\alpha = (K_{\rho}+K_{\rho}^{-1}-2)/4$
since the notion of a $K_{\sigma}$ does not make sense). Folklore would
predict another singularity  
$|\omega - \varepsilon_{\rho}(q)|^{(\alpha-1)/2}$ (dashed lines in
Fig.\ 2) 
which is \em not \rm observed here. It is cut off instead to a 
finite maximum of order $\Delta_{\sigma}^{(\alpha-1)/2}$: As in the
1D quantum antiferromagnet, the opening of the spin gap cuts off the 
singularity of the prefactor of the delta function in the spin-equivalent
to Eq.~(\ref{chargefree}) as 
$q \rightarrow 0$. \em The spin gap therefore supresses the divergence
associated with the charge dispersion while on the renormalized spin 
dispersion, the spectral response remains singular. \rm  

At negative frequencies, the Luther-Emery model has pronounced shadow bands. 
Here, the Luttinger liquid only has very small weight. 
The weight in the Luther-Emery model
is much stronger here, and the spectral function has the same overall shape
as at positive frequencies. For $q>0$, the positive frequency part is 
enhanced by a coherence factor 
$1 + v_{\sigma} q / \varepsilon_{\sigma}(q)$
while a factor $1 - v_{\sigma} q / \varepsilon_{\sigma}(q)$ decreases its
shadow. These factors translate the increased coherence due to the 
spin pairing.

Can we expect structured spectral functions for $\alpha$ larger than 1/2 or 
1? The present calculation which amounts to determining the leading behavior
does not allow a definite answer. Specifically, we have been generous on
details of cutoff procedures and therefore do not fulfill the sum rules.
Experience with the Luttinger model shows, however, that once all sum rules
are enforced, when the exponents increase so as to change a divergence into
a cusp singularity the prefactor changes sign so as to turn upward the 
cusps \cite{sanseb}. Such a crossover, keeping peaky structures also for large
$\alpha$, is natural and is expected to occur in the present problem, too.

Notice finally that the behavior of $\rho(q,\omega \approx \pm
\Delta_{\sigma})$ is determined
by that of the spin part close to $\Delta_{\sigma}$ and the charge part
at $\omega \approx 0$. It is therefore \em not \rm necessary to know details
of the charge dynamics on a scale $\omega \approx \Delta_{\sigma}$ where
the Luttinger description may have acquired significant corrrections. 

The $k$-integrated density of states then is $N(\omega) \sim
\Theta(\omega - |\Delta_{\sigma}|)  (\omega - |\Delta_{\sigma}|)^{\alpha}$
with $\alpha$  given above. There is no weight below the gap, and the
typical gap singularity in the density of states of the spin 
fluctuations is wiped out by the gapless charges.

The spectral function of a 1D Mott insulator can be computed similarly
($\sigma \leftrightarrow \rho$ everywhere). 
Spin-rotation invariance, however, requires $K_{\sigma} = 1$, and 
Eq.~(\ref{chargefree}) must be used for the spin part. 
Then $\rho(q,\omega) \propto 
\Theta(\omega - \sqrt{v_{\rho}^2 q^2 + \Delta_{\rho}^2}) /
\sqrt{\omega - \sqrt{v_{\rho}^2 q^2 + \Delta_{\rho}^2}} $,
a consequence of the convolution of two delta functions now. 
Continuity then suggests that as the Mott transition
is approached by varying the band-filling, spectral weight is
gradually taken out of both the charge and spin divergences of the Luttinger
liquid parts of the spectral function to reappear in the Luther-Emery function
possessing only a charge divergence, although the transition leaves the spins
unaffected and opens only a charge gap. $N(\omega) \sim \Theta(\omega - 
\varepsilon_{\rho})$ here. 

A wide variety of models fall into the Luther-Emery universality
class and the present results should be applicable there in a low-energy
sector: Luttinger liquids coupled to phonons and related models so long as
they are incommensurate, have wide regions of parameter
space with gapped spin fluctuations \cite{zkl}; the negative-$U$ Hubbard 
model at any band-filling has a spin gap \cite{negu}, and the positive-$U$ 
Hubbard model at half-filling has a charge gap 
\cite{liebwu,preuss}; spin gaps occur 
frequently in models of two coupled Luttinger or Hubbard chains 
\cite{twoch,tsune}, etc. Some numerical studies have attempted to
calculate spectral properties \cite{preuss,tsune}. While
consistent with the present work on the existence of shadow bands, 
their resolution is not good enough
to probe the finer structures computed here. 

Importantly, these results could prove useful in the description of the 
photoemission properties of certain quasi-1D materials. 
There is by now a considerable number of such experiments on 
quasi-1D conductors in their ``normal'' metallic state (above
low-temperature phase transitions) \cite{veuill,dardel,gweon}.
Usually, they measure the density of states
$N(\omega)$,  
which universally show an absence of
spectral weight at the Fermi edge, and a gradual raise with energy only over 
a considerable fraction of the conduction band width, these two features
being essentially temperature-independent. This behavior is 
formally consistent with the Luttinger liquid picture, predicting
$N(\omega) \propto | \omega |^{\alpha}$ with some interaction-dependent 
exponent $\alpha > 0$.  More strikingly even,
an angle-resolved photoemission experiment on $K_{0.3} Mo O_3$ shows
\em two \rm dispersing peaks \cite{gweon}.
While some materials such as the Bechgaard salts, 
may well fall into this universality
class \cite{tm}, it is particularly surprising that CDW systems 
such as the blue bronze $K_{0.3} Mo O_3$, or 
$(Ta Se_4)_2 I$ should behave similarly. In fact, the photoemission properties
are in striking contrast to the established picture of a fluctuating Peierls
insulator \cite{lra}. It predicts a strongly temperature dependent, narrow
[$| \omega | \leq \Delta_{CDW}(T=0)$] pseudogap and  $\rho(q > 0,\omega)$ is
governed by a broadened quasi-particle peak at $\omega > 0$ 
and a weak shadow at $\omega < 0$. 

A Luttinger liquid interpretation for the CDW photoemission
is highly suggestive but  encounters problems which are all resolved in
a Luther-Emery framework. (i) Luttinger 
liquids have no dominant $2k_F$-CDW correlations: for repulsive interactions
($K_{\rho} < 1$), spin density waves are logarithmically stronger than
CDWs, and for attractive interactions, the system is dominated by 
superconductivity \cite{myrev}. A spin gap is a
necessary condition for dominant CDW correlations in 1D and 
realized in the Luther-Emery model!
(ii) $2k_F$-CDWs often 
are due to electron-phonon coupling, and renormalization group
provides us with a detailed scenario
\cite{myrev,zkl}. In Fig.\ 2, we summarize the
dependence of this spin gap on electron-phonon coupling $\lambda$, the
phonon frequency $\omega_D$, and $K_{\rho}$, as calculated from
earlier results \cite{zkl}.
A spin gap also opens if CDWs are caused by Coulomb interaction between
chains \cite{iccoul}.
(iii) The spin susceptibility of CDW systems decreases with
decreasing temperature indicative of activated spin fluctuations.
(iv) For a Luttinger model, the stronger divergence in $\rho(q,\omega)$
is associated with the charge mode. For repulsive interactions, $v_{\rho}
> v_{\sigma}$ while in the experiment on $K_{0.3} Mo O_3$, the quickly
dispersing signal is less peaked than the slow one. On the other hand,
the important feature of the Luther-Emery spectral function, Fig.\ 1, 
is that the spin gap supresses the divergence of the charge signal which
disperses more quickly than the divergent spin contribution. The 
Luther-Emery spectral function is consistent with the experiments and this 
model therefore might be a natural starting point for a
description of the low-energy physics of CDW materials such as 
$K_{0.3} Mo O_3$.

Obviously, this suggestion is somewhat speculative and independent support
is called for. Its virtue is that it comes
to grips with the puzzle that the spin susceptibility of $K_{0.3} Mo O_3$
decreases with decreasing temperature while the conductivity is metallic, 
that it leaves space for the good description of optical
properties as a  fluctuating Peierls insulator (they only probe the charge
fluctuations which will form CDW precursors at temperatures much below the
spin gap opening, presumably as a consequence
of emerging 3D coherence), and that it 
provides an (admittedly phenomenological)
description of the photoemission properties of this material with extremely
1D \em electronic \rm properties \cite{pouget}. 
As in the Bechgaard salts \cite{tm}, 
a single-particle exponent 
$\alpha \sim 1$ would be required implying strong long-range
electron-electron interactions, and there is at best preliminary
support from transport measurements, for
such strong correlations in  $K_{0.3} Mo O_3$. 
Retarded electron-phonon coupling
could increase $\alpha$ over its purely electronic value \cite{zkl}.
To what extent this mechanism contributes can be gauged from the measured
$\alpha$ which must be larger than the one derived from the enhancement of
$v_{\rho}$ over the band velocity (h\'{e}las strongly depending on the
accuracy of band structure calculations).
In the perspective of the present work, high-resolution
photoemission studies on the organic conductor TTF-TCNQ are 
desirable because there is independent evidence both for strong electronic
correlations and electron-phonon coupling, and a crossover between regimes
dominated by one or the other seems to take place as the temperature is
varied.

\acknowledgements
I wish to acknowledge fruitful discussions with J. W. Allen, W. Brenig,
R. Claessen, M. Grioni, M. Gul\'{a}csi, G.-H. Gweon, D. Malterre, and J.-P.
Pouget. This research was supported by  DFG under SFB 279-B4.

\figure{FIG.~1 Spectral function of the Luther-Emery model for $q>0$.
The thick dashed line at $\varepsilon_{\rho}(k)$ gives the Luttinger liquid
divergence which is supressed here.  
The inset shows the dispersion of the two $\omega>0$-features.}

\figure{FIG.~2 Schematic dependence of the spin gap in a Luttinger liquid 
coupled
to phonons on phonon frequency for various electron-electron interactions
$K_{\rho}$ and fixed electron-phonon coupling $\lambda$.}

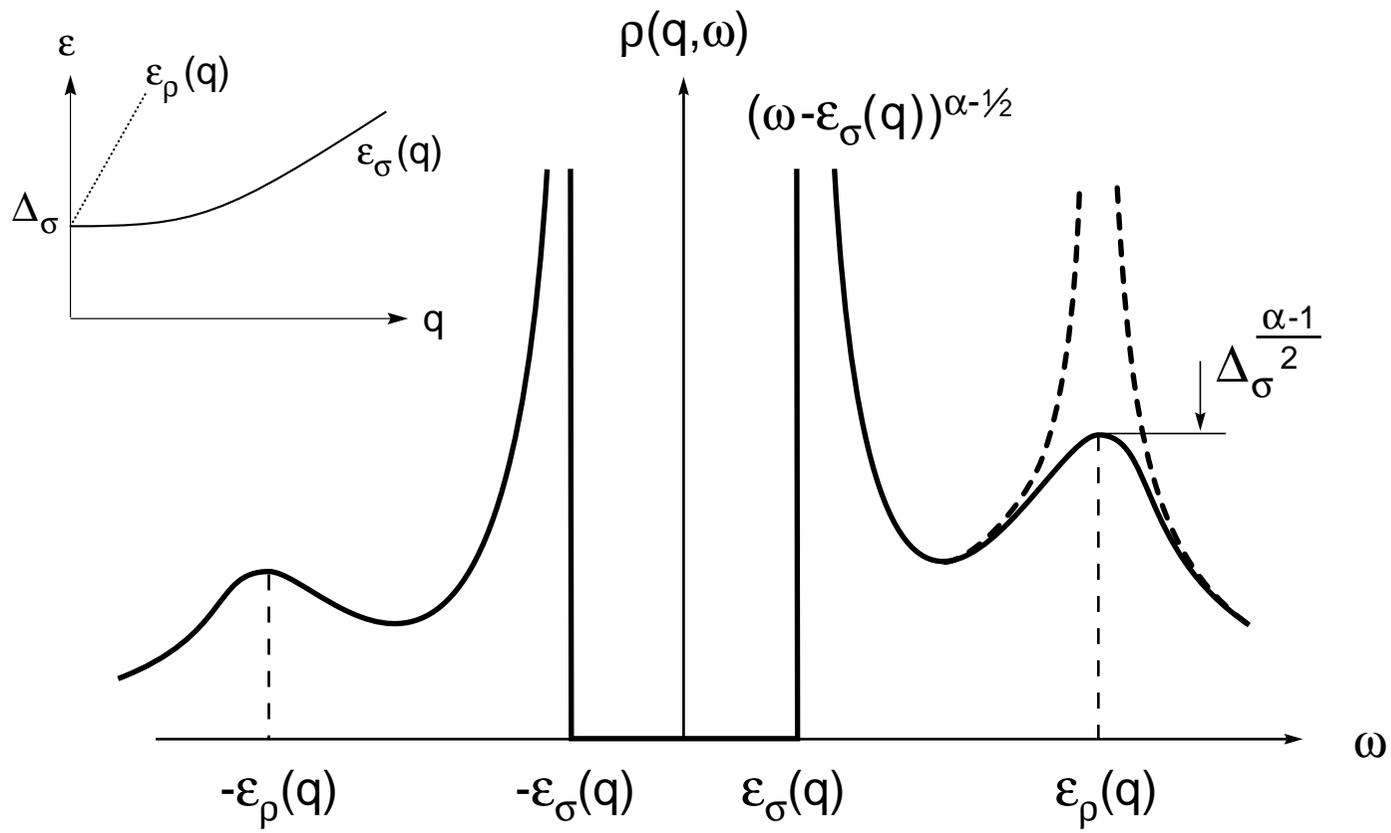
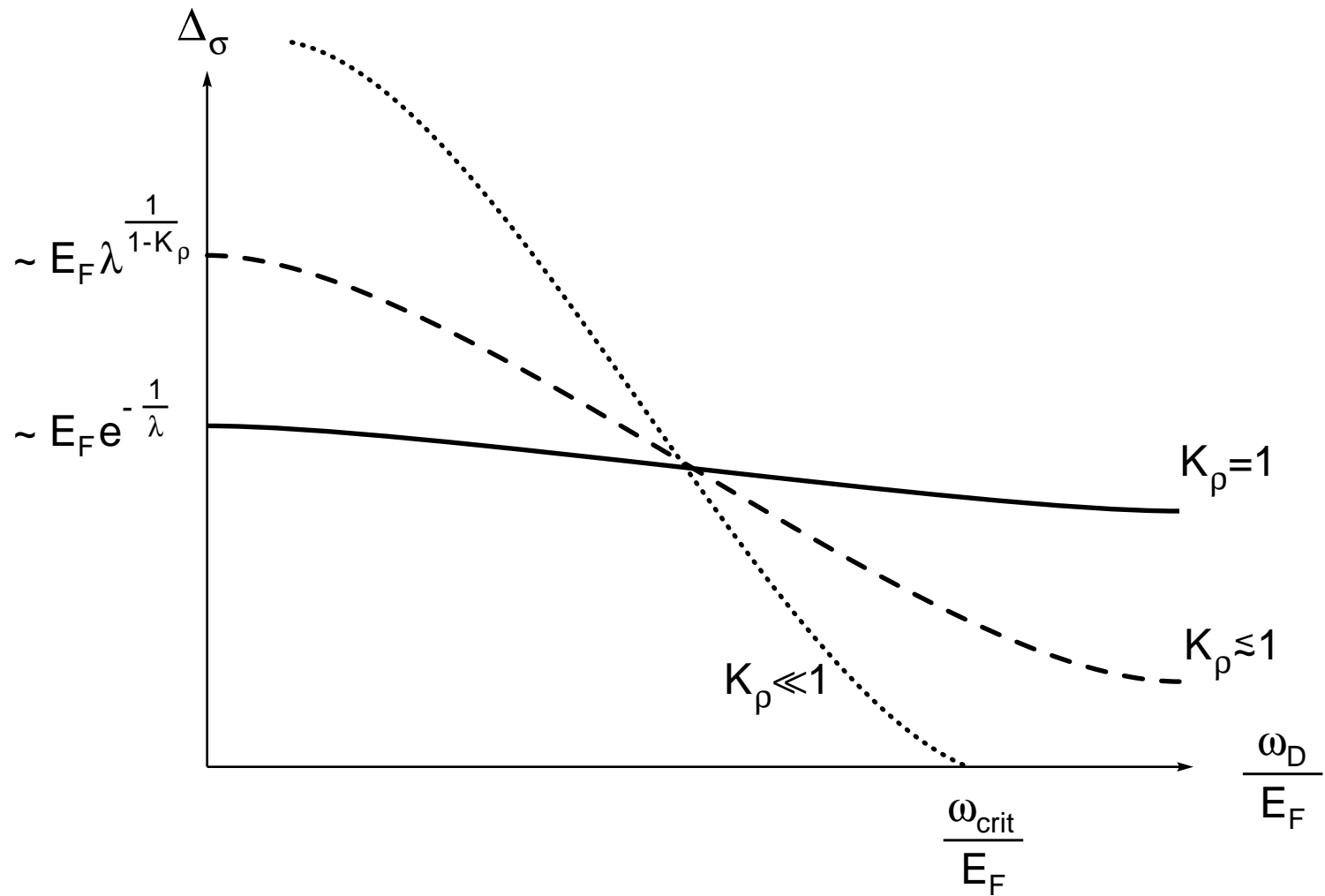
\end{document}